\documentclass[floatfix,twocolumn,amsmath,amssymb,aps,prb]{revtex4}
\usepackage{times}
\usepackage{epsfig}
\usepackage{color}
\usepackage{dsfont}

\begin{document}

\title{Green functions of interacting systems in the strongly localized regime}

\author{A. M. Somoza and M. Ortu\~no}
\affiliation{Departamento de F\'isica -- CIOyN, Universidad de Murcia, Murcia 30.071, Spain}

\author{V.\, Gasparian}
\affiliation{California State University, Bakersfield, California, USA}

\author{M.\ Pino}
\affiliation{Department of Physics and Astronomy, Rutgers The State University of New Jersey\\
136 Frelinghuysen Rd, Piscataway, 08854 New Jersey, USA}

\begin{abstract}
We have developed an approach to calculate the single-particle Green function of a one-dimensional many-body system
 in the strongly localized limit at zero temperature.
Our approach, based on the locator expansion, sums the contributions of all possible forward scattering paths in configuration space.
We demonstrate for fermions that the Green function factorizes when the system can be splited into two non interacting regions. This implies that for nearest neighbors interactions the Green function factorizes at every link connecting two sites with the same occupation. 
As a consequence we show that the conductance distribution function for interacting systems is log-normal, in the same universality class as for non-interacting systems. 
We have developed a numerical procedure to calculate the ground state and the Green function, generating all possible paths in configuration space. We compare the localization length computed with our procedure with the one obtained via exact diagonalization.  The latter smoothly converges to our results as the disorder increases.

\end{abstract}
\maketitle

\section{Introduction}

After more than 50 years of  Anderson's seminal work\cite{And58}, the problem of localization remains of current interest. Due to the complexity introduced by the crucial role of disorder, a full analytical treatment of the problem remains elusive, even for noninteracting systems.

An important idea that influenced the whole field was single-parameter scaling (SPS)\cite{AA79}. Assuming that the localization length $\xi$ is the only relevant length in the problem, it is possible to argue that one-dimensional (1D) and two-dimensional (2D) systems always flow, in the renormalization group sense, towards the strongly localized regime. The SPS assumption leads to universal results for the full conductance distribution $f(g,\xi/L)$, where $g$ is the conductance and $L$ the system size. 
This assumption becomes natural at the Anderson transition when the localization length diverges, but the same universality should be expected as long as $\xi$ is the longer microscopic length in the system. This idea has been checked analytically for 1D systems, where the distribution is log-normal and in the SPS limit\cite{shapiro} $\sigma^2=\mu$. It has also been tested numerically for several 1D and 2D systems \cite{KrMc93}. 
The idea is, thus, relevant in order to obtain the different universality classes. In this context, it is important to know if interacting systems are in the same universality class as non interacting systems. An analytical treatment of the problem is essentially impossible up to date and a numerical diagonalization study is limited to very small system sizes, so no reliable distribution function of conductance can be obtained.

In this paper we study 1D fermionic interacting systems at $T=0$ following an alternative route. In 1D  the localization length itself is an irrelevant variable, in the renormalization group sense, as the system flows to the fixed point $\xi/L \rightarrow 0$. We can expect that universality emerges as we approach this fixed point.  In order to obtain the corresponding universal distribution we can work with a system very near this fixed point. This procedure has been already applied to non interacting systems with great success.

Starting with an Anderson model, it is possible to calculate the Green function (GF) via a locator expansion and keep only the lowest order terms (which correspond to considering only forward-scattering paths). This type of approach was used by Nguyen, Spivack and Shklovskii \cite{NS85} for single-particle propagation, although they were mainly interested in studying the magnetoresistance and not the conductance distribution function.   It was also applied to 2D systems by Medina and Kardar \cite{MK92}, who showed that the variance of $\log g$ grows with system size as $L^{2/3}$, result which was later confirmed  for the Anderson model \cite{PS05}. 
It is easy to show that, for 1D systems, this approach reproduces in a naive way the conductance log-normal distribution. For 2D systems, we were able to show that the universal conductance distribution function is a Tracy-Widom distribution that
depends on boundary conditions. We analyzed the cases of narrow and wide leads \cite{SoOr07,PSO09} and also a half-plane system\cite{SDO15}. The applicability of these universal distributions to the Anderson model in the strongly localized regime was confirmed in all cases.

In this paper we study the localized regime of an interacting system in 1D following the forward-scattering paths approach. 
In the next section we present the model employed, and in Sec.\ \ref{paths} we develop a procedure to calculate the leading contribution to  the single-particle GF of the many-body system in terms of forward-scattering paths in configuration space. We then prove, in section \ref{factor}, that the GF can be factorized
when there is no effective interaction in a link of the system. 
We present numerical results in \ref{numeric} and compare our predictions with results obtained with exact diagonalization for small system sizes. In \ref{Sapprox} we present a practical way to implement our method approximately. We end up this paper with a discussion of the results and extracting some conclusions.

\section{Model}

We are interested in calculating the d.c. conductance, $g$, for a localized system at zero temperature. It can be obtained from the Landauer formula\cite{Datta} $g={\rm tr}\{ \Gamma G_{\rm R} \Gamma G_{\rm A}\}$, in units of $e^2/h$, where $G_{\rm R}$
($G_{\rm A}$) is the retarded (advanced) Green function and $\Gamma$ the coupling of the system to the lead. 
This expression was deduced for noninteracting systems, but  remains valid for interacting systems at zero temperature \cite{MW92}. $G_{\rm R} = G_{\rm R}({\bf r},{\bf r}',\omega)$ is the Fourier transform of the time dependent one-particle retarded Green function, defined as\cite{Abrikosov}
\begin{equation}
G_{\rm R}(x,t;y,t')=\begin{cases}-i\langle \tilde{c}_{y}(t') \tilde{c}_{x}^\dagger(t)+ \tilde{c}_{x}^\dagger(t) \tilde{c}_{y}(t')\rangle 
&\mbox{if } t'>t \\
0&\mbox{if } t'<t \end{cases}
\label{retarded}
\end{equation}
where $\tilde{c}_{x}(t)$ and $\tilde{c}_{x}^\dagger(t)$ are destruction and creator operators, respectively, of a particle at position $x$ in time $t$  in Heisenberg representation, and $\langle\cdots\rangle$ refers to expected values in a given state, which in this work we usually consider to be the
ground state $|\Psi_{\rm gs}\rangle$. So, we are interested in
\begin{eqnarray}
G_{\rm R}(x,y;\omega)=\sum_\alpha \frac{\langle\Psi_{\rm gs}^{(N)}|c_{y}|\Psi_\alpha^{(N+1)}\rangle \langle\Psi_\alpha^{(N+1)} 
|c_x^\dagger| \Psi_{\rm gs}^{(N)}\rangle}{\omega-E_\alpha^{(N+1)}+E_{\rm gs}^{(N)}+i\delta}\nonumber\\
+\sum_\beta \frac{\langle\Psi_{\rm gs}^{(N)}|c_{x}^\dagger|\Psi_\beta^{(N-1)}\rangle \langle\Psi_\beta^{(N-1)} 
|c_{y}| \Psi_{\rm gs}^{(N)}\rangle}{\omega-E_{\rm gs}^{(N)}+E_\beta^{(N-1)}+i\delta}\,.
\label{omega}\end{eqnarray}

We assume that we are dealing with an $N$ particle system, and include the number of particles of each quantity as a superscript just for clarity. $| \Psi_{\rm gs}^{(N)}\rangle$ and $E_{\rm gs}^{(N)}$ are the ground state energy and wave function, respectively. Here $c_x$ ($c_x^\dagger$) annihilates (creates) a particle at $x$. The sum over $\alpha$ ($\beta$) runs over all $N+1$ ($N-1$) particle states, and $E_\alpha^{(N+1)}$ ($E_\beta^{(N-1)}$) are their corresponding energies. We say that the first term in the RHS of (\ref{omega}) is the electron contribution to the GF, while the second term is the hole contribution.

For a system of $N$ particles, we can define the $N$-particle Green function as
\begin{equation}
G^{(N)}(\{x_i\},t;\{y_j\}, t')= -i \langle 0|\prod_{j=1}^N \tilde{c}_{y_j}(t')\prod_{i=1}^N \tilde{c}^\dagger_{x_i}(t) |0\rangle\,
\label{one1}\end{equation}
whose time Fourier transform is related to the $N$-particle Hamiltonian
\begin{equation}
G^{(N)}(\{x_i\};\{y_j\};\omega)= \left[\omega \mathds{1} -H^{(N)}\right]^{-1}.
\label{one}\end{equation}

From now on, we will restrict to 1D systems of interacting spinless fermions with a strong disorder, although most of the ideas
can be easily extended to boson systems and to higher dimensions.

We define a discrete tight binding  Hamiltonian
\begin{equation}
H^{(N)}=\sum_{x=1}^L\epsilon_x c_x^\dagger c_x -t  (c_x^\dagger c_{x+1} + c_{x+1}^\dagger c_x)
+ U c_x^\dagger c_x c_{x+1}^\dagger c_{x+1}
\label{hamil}\end{equation}
where $\epsilon_x$ is the  external random potential at site $x$, uniformly distributed in the range $[-W/2,W/2]$, $t$ is the transfer energy, $U$ the interaction energy, $c_x$ and $c_x^\dagger$ the annihilation and creation operators at site $x$. We concentrate in the strongly localized regime where $U, W \gg t$.

We consider a many-body locator expansion $H^{(N)}=H_0^{(N)}+\Delta H^{(N)}$ where $\Delta H$ is the second (hopping) term in
the RHS of (\ref{hamil}) and $H_0^{(N)}$ is a classical interacting Hamiltonian. 
Then the $N$-particle Green function obeys a Dyson equation
\begin{eqnarray}
G^{(N)}= G_0^{(N)}&+&G_0^{(N)} \Delta H^{(N)} G^{(N)}\nonumber\\
= G_0^{(N)}&+& G_0^{(N)} \Delta H^{(N)} G_0^{(N)}\\
&  +& G_0^{(N)} \Delta H^{(N)} G_0^{(N)} \Delta H^{(N)} G_0^{(N)}+ \dots\nonumber 
\end{eqnarray}
As $G_0$ is diagonal
\[
G^{(N)}_0(\{x_i\};\{y_j\}, \omega)=\frac{\delta_{\{x_i\},\{y_j\}}}{\omega-E_0(\{x_i\})}
\]
$G^{(N)}$ turns out to be the sum of all possible paths in Hilbert space from point $\{x_i\}$  to point $\{y_i\}$.

The single particle Green function, Eq.\ (\ref{omega}) can be written in terms of $G^{(N+1)}$ and $G^{(N-1)}$ 
\begin{align}
G_{\rm R}(x,y;\omega)=&\langle\Psi_{\rm gs}^{(N)}|c_{y} G^{(N+1)}(\omega+E_{\rm gs})
c_x^\dagger| \Psi_{\rm gs}^{(N)}\rangle\\
&-\langle\Psi_{\rm gs}^{(N)}|c_{x}^\dagger G^{(N-1)}(-\omega+E_{\rm gs})
c_y| \Psi_{\rm gs}^{(N)}\rangle\,.\nonumber
\label{omega2}\end{align}
Due to the presence of the wave function  $| \Psi_{\rm gs}^{(N)}\rangle$, the representation of the single particle Green function in terms of paths is not straight forward, but it is feasible in the strongly localized regime in which we are interested.

Since we are interested only in the many-body single-particle GF, from now on we will drop the subscript ``R'' referring to the retarded GF. We will take for granted that we will always refer to this GF, unless explicitly mentioned.

\section{Directed Path approach}
\label{paths}

The representation of the single-particle Green function in terms of paths is much simpler in the limit $t\rightarrow 0$. 
We will calculate $G$ up to lowest order in $t$. 
For simplicity, we restrict the discussion to 1D systems, although it is straightforward to generalize the ideas to any dimension. 
In this approach only paths in configuration space where particles move from left to right, from an occupied to empty site, are considered.
The  wave function $| \Psi_{\rm gs}^{(N)}\rangle$  includes the classical ground state configuration (to zero order in $t$) plus additional contributions to all orders in $t$, which have to be properly taken into account.
As particles are only allowed to hop to the right, the propagation process can be thought in a simple  way.
The initial and the final configurations correspond to the classical ground state, the incoming particle on the left must travel up to the position of the first particle of the system, which in turn will end up in the position of the second one, and so on. The last particle will have to move to the right end of the system, where it will be removed. 
Insertion of a particle can take place in configurations with the first site empty, while removal can only occur when the final site is occupied. 
All hops, including the insertion and extraction of a particle, can take place in any
possible order and the contributions from all paths in configuration space must be summed up to get the final GF. 
This arbitrariness in the order of the hops includes processes in which a particle is first removed from the final site and later inserted in the initial site, i.e. hole propagation.

The rules to calculate the contribution of each path are the following. 
Each time that an electron hopes corresponds to a factor $t$ and each time that a configuration $\alpha$ is visited to a factor
\begin{equation}
\frac{1}{\eta\omega +E_{\rm gs}^{(N)}-E_\alpha^{(N+\eta)}}
\label{contri}\end{equation}
where $\eta$ is the difference in the number of particles of configuration $\alpha$ and the initial configuration (usually the ground state). 
Terms with $\eta=1$ ($\eta=-1$) come from particle (hole) propagation, i.e. the expansion of $G^{(N+1)}$ ($G^{(N-1)}$) in the first (second) term in the RHS of (\ref{omega}).
The terms with $\eta=0$ arise from the expansion of  $| \Psi_{\rm gs}^{(N)}\rangle$ in $t$, and can be viewed as hops occurring
before the particle is inserted and after it is removed. 
Along any of the paths, the final result is that a particle has traveled from left to right or a hole from right to left, with the number of steps equal to the number of links between sites, after which the system returns to its original state.
The energy of a configuration is the sum of the disorder site energies $\epsilon_x$ of all occupied places plus an interaction energy $U$ for each pair of occupied nearest neighbor sites.

In the absence of interactions, it is possible to prove that the sum of the contributions from all paths is equal to the expression of the one-particle GF, which in the strongly localized limit consists of a single term.  Interactions modify the relative contributions of the different paths resulting in complex, although well controlled expressions, composed by more than one term.

An example of the implementation of the procedure for a system with two sites is shown in  the next subsection,
and a more complex case with three sites is shown in  appendix A.

\subsection{Two sites}

\begin{figure}
\includegraphics[width=.35\textwidth]{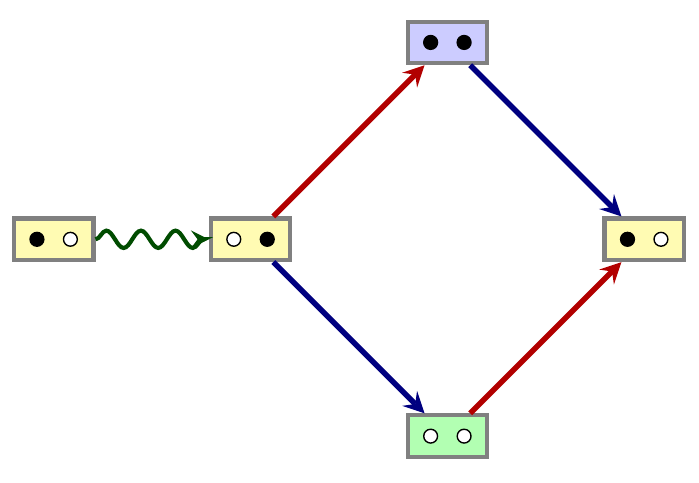}
\caption{Scheme of the configurations involved in the calculation of the GF for a system with two sites and one electron, originally placed at the first site. Each rectangle corresponds to a configuration. The red arrow marks the insertion
of an electron at site 1, while the blue arrow signals the extraction of an electron at site 2, and the wavy arrow propagation by the hopping $t$.
\label{two}}
\end{figure}

Let us consider the simplest system consisting of just two sites and one electron. For the sake of concreteness, let us assume that in the ground state, to zero order in $t$, the first site is occupied and the second one empty.
In the noninteracting case the GF between sites 1 and 2 must be equal (up to a sign) to the one-particle GF, defined as 
in Eqs.\ (\ref{one1}-\ref{one}) with $N=1$.
In the strongly localized limit, the latter is 
\begin{equation}
G^{(1)}(1,2;\omega)=\frac{1}{\omega-\epsilon_1}t\frac{1}{\omega-\epsilon_2}\,.
\label{app1}\end{equation}
We first recover this result with our path approach in configuration space. In this case, there are only two paths, schematically represented in  Fig.\ \ref{two}, where each rectangle corresponds to a configuration and the lines are the possible transitions between configurations. Upwards lines represent the creation of a particle at the initial site 1, and downwards lines the annihilation of a particle at $L=2$. The wavy line corresponds to a hop from one site to an adjacent one and contributes with a factor $t$. The two path start with the jump of the particle at 1 to 2. Then, one path consists in the injection of a particle at 1 and the subsequent annihilation of a particle at 2, while in the other path the particle at 2 is annihilated first and a new particle is created at 1. Their contributions are
\begin{equation}
G(1,2;\omega)=\frac{t}{\epsilon_1-\epsilon_2}\left[
\frac{1}{\omega-\epsilon_2}+\frac{1}{-\omega+\epsilon_1}\right]
\label{app2}\end{equation}
and is trivial to check that the sum coincides with Eq.\ (\ref{app1}), up to a sign.

In the presence of interactions, we have to add the interaction energy $U$ to the configuration where both sites are occupied. 
Then Eq.\ (\ref{app2}) becomes
\begin{equation}
G(1,2;\omega)=t\frac{1-U/(\epsilon_1-\epsilon_2)}{(\omega-\epsilon_2-U)(-\omega+\epsilon_1)}
\label{app13}\end{equation}

In this simple case, we can check the validity of our path approach with the spectral representation of the GF, Eq.\ (\ref{omega}), to lowest order in $t$.

\subsection{Three sites}

As the system becomes larger, the number of paths involved increases drastically, but it is still manageable up to
relatively large sizes, while a direct evaluation of the spectral representation becomes impractical.
A further example of our method for a system of three sites (in an initial state with one particle at the site in the middle) is shown in appendix \ref{appa}.
The GF for the interacting case is given by (\ref{Ga})
\begin{equation}
G(1,3)=\frac{1-\frac{U}{\epsilon_2-\epsilon_1}-\frac{U}{\epsilon_2
-\epsilon_3}+
\frac{U^2}{w+\epsilon_2-\epsilon_1-\epsilon_3}(\frac{1}{
\epsilon_2-\epsilon_1}+\frac{1}
{\epsilon_2-\epsilon_3})}
{(w-\epsilon_1-U)(w-\epsilon_2)(w-\epsilon_3-U)}.
\label{tres}\end{equation}
Here we have taken $U_1=U_3=U$ for simplicity.
In appendix \ref{appa} we show how this expression can be obtained also from the spectral representation of the GF to lowest order in $t$.

The different terms in (\ref{tres})  take into account  electron correlations in the different possible paths and 
their complexity rapidly increases with system size, but we will see in the next section that under certain relevant circumstances 
the GF can be factorized.

\section{Factorization of the Green functions}
\label{factor}

\begin{figure}
\includegraphics[width=.49\textwidth]{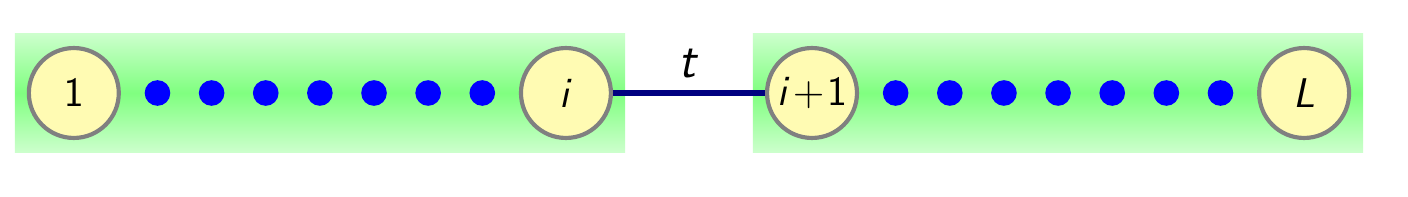}
\caption{Scheme of a system that can be divided into two subsystems with no interaction between them
and a small transfer integral $t$ between them.
\label{fact}}
\end{figure}

Let us now consider a 1D system as the one represented in Fig.\ \ref{fact}.
We assume that there are no interactions between any site to the left of $i$ (this site included)
and any site to the right of $i+1$ (this site also included).
We further suppose that the transfer energy $t$ between sites $i$ and $i+1$ is much smaller than the
other energies of the problem and are then interested in the calculation of the GF 
$G(1,L;\omega)$ to first order in $t$.
In this section, we are going to expand the GF in terms of this particular $t$ and the rest of transfer energies
can take any value.
We also  remark that the two subsystems, from 1 to $i$ and from $i+1$ to $L$, can have any interaction
and do not have to necessarily be in the strongly localized regime.

The system is in any eigenstate $|\Psi\rangle$ of the total Hamiltonian, not necessarily the ground state.
At zero order in $t$, this state can always be written as
\begin{equation}
|\Psi_0\rangle=|\psi_0^{(N_L)}\rangle \otimes |\varphi_0^{(N_R)}\rangle
\label{fac1}\end{equation}
$N_L$ ($N_R$) indicates the number of particles in the left (right) subsystem, at zero order in $t$, 
and the total number of particles is $N=N_L+N_R$.
We use the symbol $\psi$ ($\varphi$) to refer to wave functions on the left (right). 
The total energy is $E_0=E_0^{(N_L)}+E_0^{(N_R)}$.
The first order contribution to $|\Psi\rangle$ is
\begin{eqnarray}
&&|\Delta\Psi\rangle=t\sum_{\alpha,\beta}\frac{\langle\psi_\alpha^{(N_L-1)}|c_i|\psi_0^{(N_L)}\rangle
\langle\varphi_\alpha^{(N_R+1)}|c_{i+1}^\dagger|\varphi_0^{(N_R)}\rangle}{E_0^{(N_L)}+E_0^{(N_R)}-E_\alpha^{(N_L-1)}-E_\beta^{(N_R+1)}}\nonumber\\
&&|\psi_\alpha^{(N_L-1)}\rangle \otimes |\varphi_\beta^{(N_R+1)}\rangle
\label{fac2}\end{eqnarray}
The sum in $\alpha$ ($\beta$) runs over all eigenstates in the left (right) side with $N_L-1$ ($N_R-1$) particles.
The previous expression corresponds to a particle moving from left to right.
There is a similar expression for a particle crossing in the opposite direction, but it does not contribute to the GF to lowest order in $t$.

We have to evaluate (\ref{omega}) to first order in $t$ for our present system.
The final result is a sum of six terms, three coming from the particle contribution to the GF, $G^{(N+1)}(\omega +E_0^{(N)})$, and the other three from the hole contribution, $G^{(N-1)}(-\omega +E_0^{(N)})$. The three terms of each set correspond to linear contributions from the bra wave function, from the GF and from the ket wave function, respectively. We have
\begin{eqnarray}
G(1,L;\omega)&=&
\langle\Psi_0| \Delta G^{(N+1)}(\omega +E_0^{(N)})|\Psi_0\rangle\nonumber\\
&+&\langle\Psi_0|  \Delta G^{(N-1)}(-\omega +E_0^{(N)})|\Psi_0\rangle\nonumber\\
&+&\langle\Delta\Psi| G^{(N+1)}(\omega +E_0^{(N)}) |\Psi_0\rangle\nonumber\\
&+&\langle\Delta\Psi| G^{(N-1)}(-\omega +E_0^{(N)}) |\Psi_0\rangle\nonumber\\
&+&\langle\Psi_0| G^{(N+1)}(\omega +E_0^{(N)}) |\Psi_0\rangle\nonumber\\
&+&\langle\Psi_0| G^{(N-1)}(-\omega +E_0^{(N)}) |\Psi_0\rangle
\label{fac3}\end{eqnarray}

The first term on the RHS of (\ref{fac3}) can be written as
\begin{eqnarray}
{\rm GF}(1)=\langle\psi_0^{(N_L)}| \langle\varphi_0^{(N_R)}|
c_L G_0^{(N+1)}(\omega +E_0^{(N)})\nonumber\\ 
\Delta H G_0^{(N+1)}(\omega +E_0^{(N)}) c_1^\dagger
|\psi_0^{(N_L)}\rangle |\varphi_0^{(N_R)}\rangle
\label{fac4}\end{eqnarray}
where $\Delta H=-t c_{i+1}^\dagger c_i$. 
We can now use the spectral representation for $G_0^{(N+1)}$, Eq.\ (\ref{omega}),
taking into account that the second GF in (\ref{fac4})  must have $N_L+1$ electrons
on the left  and $N_R$ on the right. Analogously, the first GF in (\ref{fac4})  must 
have $N_L$ electrons on the left  and $N_R+1$ on the right. Thus
\begin{eqnarray}
{\rm GF}(1)=\sum_{\alpha,\beta} \frac{\langle\varphi_0^{(N_R)}|c_{L}|\varphi_\beta^{(N_R+1)}\rangle \langle\psi_0^{(N_L)}|\langle\varphi_\beta^{(N_R)}|}
{\omega+E_0^{(N)}-E_0^{(N_L)}-E_\beta^{(N_R+1)}+i\delta}\nonumber\\
\Delta H\frac{|\psi_\alpha^{(N_L+1)}\rangle |\varphi_0^{(N_R)}\rangle\psi_\alpha^{(N_L+1)}|c_{1}^\dagger|\psi_0^{(N_L)}\rangle} {\omega+E_0^{(N)}-E_\alpha^{(N_L+1)}-E_0^{(N_R)}+i\delta}\,.
\label{fac5}\end{eqnarray}
and remembering that  $E_0^{(N)}=E_0^{(N_L)}+E_0^{(N_R)}$, 
we get
\begin{eqnarray}
{\rm GF}(1)=\sum_{\alpha} \frac{ \langle\psi_0^{(N_L)}|c_i |\psi_\alpha^{(N_L+1)}\rangle
\langle\psi_\alpha^{(N_L+1)}|c_{1}^\dagger|\psi_0^{(N_L)}\rangle}
{\omega+E_0^{(N_L)}-E_\alpha^{(N_L+1)}+i\delta}\nonumber\\
t\sum_{\beta}\frac{\langle\varphi_0^{(N_R)}|c_{L}|\varphi_\beta^{(N_R+1)}\rangle
\langle\psi_\beta^{(N_R+1)}|c_{i+1}^\dagger|\psi_0^{(N_R)}\rangle } {\omega+E_0^{(N_R)}-E_\beta^{(N_R+1)}+i\delta} \,.
\label{fac6}\end{eqnarray}
The first sum in (\ref{fac6}) corresponds precisely to the electron contribution to the GF between 1 and $i$, while the second sum corresponds to the electron contribution between $i+1$ and $L$. 
We finally arrive at
\begin{equation}
{\rm GF}(1)=G_{\rm el}(1,i;\omega)tG_{\rm el}(i+1,L;\omega) .
\label{fac7}\end{equation}
A similar procedure for the second term on the RHS of (\ref{fac3}) produces the hole-hole contribution
\begin{equation}
{\rm GF}(2)=G_{\rm ho}(1,i;\omega)tG_{\rm ho}(i+1,L;\omega) .
\label{fac8}\end{equation}
The combined contribution of terms 3 and 4 on the RHS of (\ref{fac3}) and that of  5 and 6 result, respectively in
\begin{eqnarray}
{\rm GF}(3)+{\rm GF}(4)&=&G_{\rm el}(1,i;\omega)tG_{\rm ho}(i+1,L;\omega)\\
 {\rm GF}(5)+{\rm GF}(6)&=&G_{\rm ho}(1,i;\omega)tG_{\rm el}(i+1,L;\omega).
\label{fac9}\end{eqnarray}
Adding together all six contributions (\ref{fac7}-\ref{fac9}), the final result is
\begin{equation}
G(1,L;\omega)=G(1,i;\omega)tG(i+1,L;\omega) .
\label{fac10}\end{equation}

This factorization of the GF is a central result of the paper. It is a rather general result only requiring  a large tunnel barrier between two regions and weak interactions between them. 
It may be applied to different materials of technological importance such as polymer melts and granular metals. We note that this factorization is not verified by other apparently simpler quantities like, for example, the correlation function
\begin{equation}
c(1,L)= \langle\psi_0^{(N)}|c_{L} c_{1}^\dagger|\psi_0^{(N)}\rangle.
\end{equation}

The factorization is also specially relevant from a practical point of view in our calculations on the strong localization limit even in the presence of interactions.
Whenever the classical ground state has two nearest neighbors with the same occupancy the GF factorizes.
When two adjacent sites are empty (at zero order in $t$), as particles only move to the right, they are never both occupied in any of the configurations relevant for the GF and so there is no effective interaction between these sites. Then the GF factorizes as a product of the GF from 1 to the first of these two adjacent sites times the GF from the second site to $L$.
The same argument applies to two consecutive occupied sites by a symmetry argument.
To show this explicitly, we rewrite the corresponding interaction between these two sites as
   \begin{equation}
Un_i n_{i+1}= U(1-n_i) (1-n_{i+1})+U(n_i +n_{i+1})-U,
\label{fac11}\end{equation}
where $n_i=c_i^\dagger c_i$. The first term on the RHS of (\ref{fac11}) is the interaction between two empty sites (which never takes place in our approach).
The second term is an effective local potential $U$ on each of the two sites involved that can be added to the disorder energy, and the third term is just a constant.
The GF factorizes again, but we have to take into account that the effective disorder energy of each site is increased by the amount $U$.

As the GF can be factorized whenever two adjacent sites are either both occupied or both empty, the hard part of the problem is just the  calculation of  the GF of chains with alternating occupancy.

\section{Numerical  results}
\label{numeric}
\subsection{Methods}

We have implemented a numerical procedure to evaluate the GF for the ground state in the strongly localized limit.
The algorithm is as follows.
For each realization of the site disorder energies, we first calculate the exact classical ground state (in the absence of the transfer energy)
at constant chemical potential $\mu$.
We calculate the ground state occupation with an iterative procedure in the spirit of the transfer matrix technique. 
Starting form the left end of the sample, we advance site by site keeping the configurations with lowest free energy for the two possible 
occupations of the last site. Then we consider a new site and update those configurations. 
 When there is a region where the two sequences coincide, we know that this common sequence is that of the ground state. 
When we reach the right end of the system we select the  appropriate sequence according to the boundary condition.
We concentrate on half occupation ($\mu=U$) and we add a boundary term in the Hamiltonian, 
$H_{\rm b}=(c_1^\dagger c_1 + c_L^\dagger c_L)U/2$, to preserve this occupation near the borders \cite{mackinnon}.
Once we have the classical ground state we look for all the pairs of adjacent sites with the same occupation, where we know that the GF factorizes.
For each chain of alternating occupations we obtain the GF summing the contributions of all possible paths where particles jump to the right. 
We study how the GF  varies as a function of the ratio $U/W$. 
 The only limiting factor of the procedure is the probability of finding a long chain with alternating occupancies in the ground state, since the number of configurations involved grows exponentially with chain length. 

In the strongly localized regime, $\ln g$ is the self-averaging quantity, and as $g\propto |G_{\rm R}|^2$ we average $\langle \ln |G(1,L;\omega)|\rangle $ over disorder realizations. 
To an excellent approximation this average  is proportional to system size and related to the  localization length $\xi$ by
\begin{equation}
-\frac{1}{\xi}=\frac{\ln |G(1,L;\omega=\mu)|}{L}
\label{17} 
\end{equation}
For sizes up to $L=40$ we can handle all the values of the ratio $U/W$.

We also employ exact diagonalization of small systems in order to compare with our results.
The localization length is obtained by measuring the sensitivity of the ground-state energy under a change in the boundary conditions. 
Intuitively, only particles with certain probability of being near the edges of the system will notice these conditions.
Then, the sensitivity of physical quantities to the boundary terms should decay exponentially in size for localized systems.
We expect for a one-dimensional system\ \cite{EdTh72,KrMc93}:   
\begin{align}\label{Eq:Loc_length}
\lim_{L \to \infty}\ln{\left(L\ \Delta E\right)}= -\frac{L}{\xi}+c.
\end{align}
where $c$ is a constant and $\Delta E=\left| E_{\rm gs}(\pi)-E_{\rm gs}(0) \right|$. 
The quantity $E_{\rm gs}(\varphi)$ is the ground-state energy with boundary conditions $c_{L+1}=e^{i\varphi}c_1$. 
We notice that these type of boundary conditions  occurs for particles placed in a one-dimensional ring which is
threaded by a magnetic flux $\Phi=\varphi\Phi_0/(4\pi)$, where $\Phi_0=2e/\hbar$ is the magnetic flux quantum.

We proceed as follows to compute the localization length. The energy difference between the ground-state for periodic and anti-periodic
boundary conditions, $\Delta E(L)$, is obtained for several sizes $L$ (from 8 to 22). Then, an average over many 
samples is performed to obtain  $\langle L\Delta E(L)\rangle$. We fit these data to a straight line $aL+b$. Taking into account Eq.\ (\ref{Eq:Loc_length}), 
we extract the localization length as $\xi=1/a$. 
Only the ground-state energy is needed in our computations.
For this reason, we have employed ARPACK\ \cite{LeSo98} libraries which are specially suited to compute a few eigenvalues of a sparse matrix.

\subsection{Results}

Within our approach, and using Eq.\ (\ref{17}), it is easy to show that
\begin{equation}
\frac{1}{\xi}=\frac{1}{\xi_0}+f\left(\frac{U}{W}\right)
\label{19} 
\end{equation}
where $\xi_0$ is the non-interacting localization length at a given disorder, and in the strongly localized limit  $\xi_0^{-1}=\ln(W/2t)-1$.
We have performed a systematic study of the localization length for a system of size $L=40$ at half occupation and as a function of the ratio $U/W$. We consider an energy equal to the chemical potential, $\omega=\mu$.

\begin{figure}
\includegraphics[width=.45\textwidth]{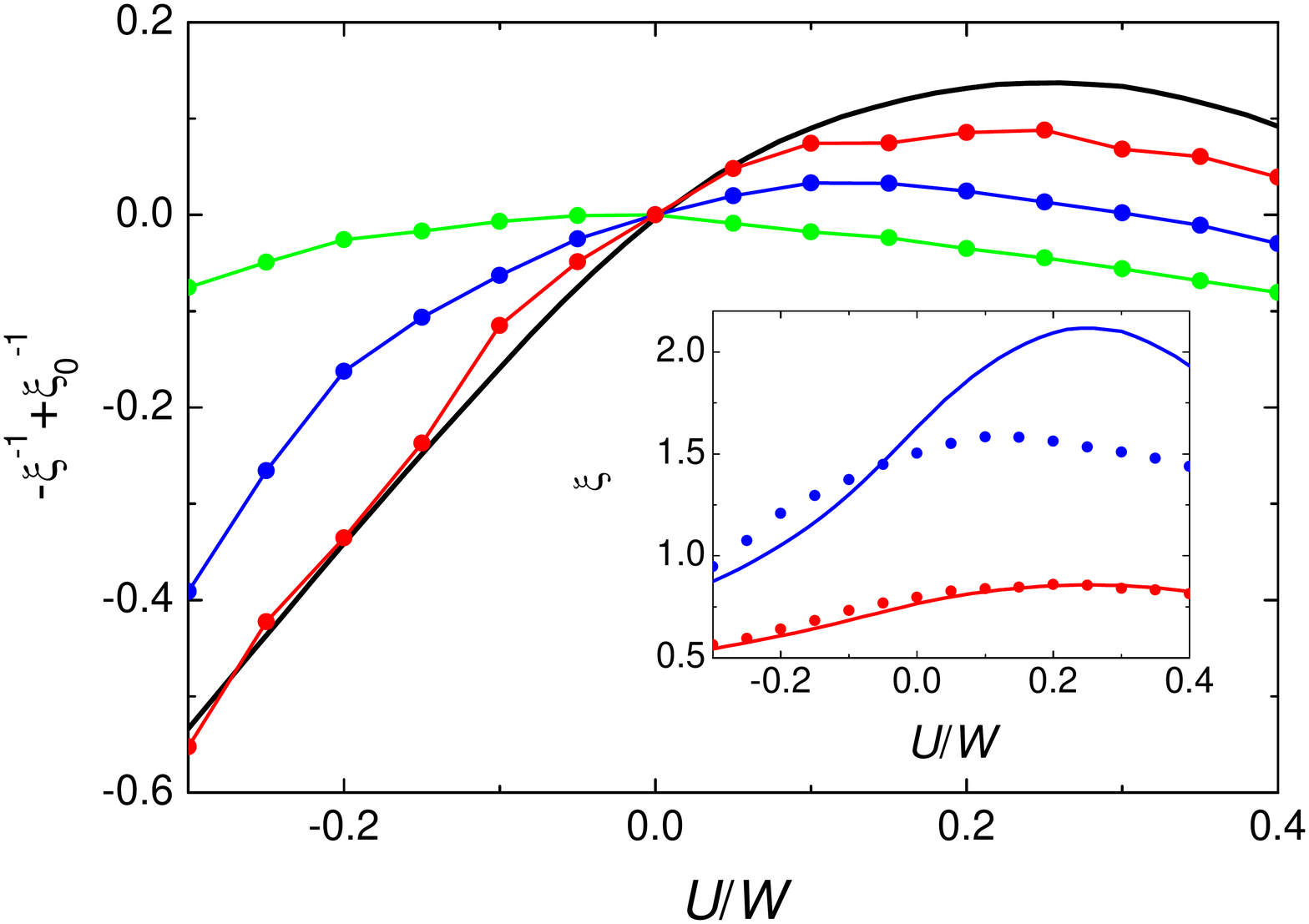}
\caption{Change in the inverse localization length with respect to the noninteracting case as a function of $U/W$ for our procedure (black curve) and for three values of the disorder $W=5$ (green), 10 (blue) and 20 (red) with exact diagonalization. The inset shows the localization length itself for the exact result (circles) and our approach (continuous curves) for $W=10$ (blue) and 20 (red).
\label{overlap}}
\end{figure}

In the inset of Fig.\ \ref{overlap} we represent the localization length versus $U/W$ for $W=10$ (blue) and 20 (red).
The symbols correspond to the exact diagonalization results and the continuous curves to our approach.
Our results can be scaled into a single curve by plotting
 $-\xi^{-1}+\xi_0^{-1}$ versus $U/W$, as it is done in the main panel of  Fig.\ \ref{overlap}. The black curve corresponds to our results, while the symbols to  exact diagonalization with $W=20$ (red), 10 (blue) and 5 (green). 
For an attractive interaction, $U<0$, the localization length decreases, with respect to the noninteracting case, while
for weak repulsive interactions, the system becomes less localized.
This is in contrast with previous simulations at moderate disorder \cite{mackinnon,ScSc98} and in agreement with  calculations for strong disorder \cite{vojta,ScJa98}.
We have  included results from exact diagonalization for large disorder in order to visualize the evolution of $\xi$
from moderate to large disorder.
As disorder increases, the exact results approach ours and show similar behavior
presenting a maximum of the localization length at $U/W$ approaching the value 0.3.

The smooth tendency of the exact results towards the $t\rightarrow 0$ limit
indicates that the perturbative  approach is reasonable.
From a quantitative point of view, results can be improved by properly modeling the effects of
 resonances (see section \ref{dis}).

\section{Factorization approximation}
\label{Sapprox}

According to the previous result on factorization, we can construct the total GF of a long chain from the expressions of the GF of short chains with alternating occupancies in the state for which we want to calculate the GF. Eqs.\ (\ref{app13}) and (\ref{tres}) are the simplest examples of such type of expressions.

As the length of the alternating sequence increases the complexity of the GF increases and it is not possible to deduce their influence in overall quantities, such as the localization length. We want to investigate the influence of such complexity on the overall GF comparing the exact results with an approximation based on factorization.

As a first step towards a factorized expression to approximate the GF we consider by analogy with the noninteracting case -in the strongly localized limit
\begin{equation}
G(1,L;\omega)\approx t^{L-1} \prod_{i=1}^{L} G(i,i;\omega).
\label{first}
\end{equation}
This expression is equivalent to the single-particle GF, but $G(i,i;\omega)$ is calculated taking into account the interaction contribution
from the rest of the system assuming these particles are frozen in the ground state.

\begin{figure}
\includegraphics[width=.45\textwidth]{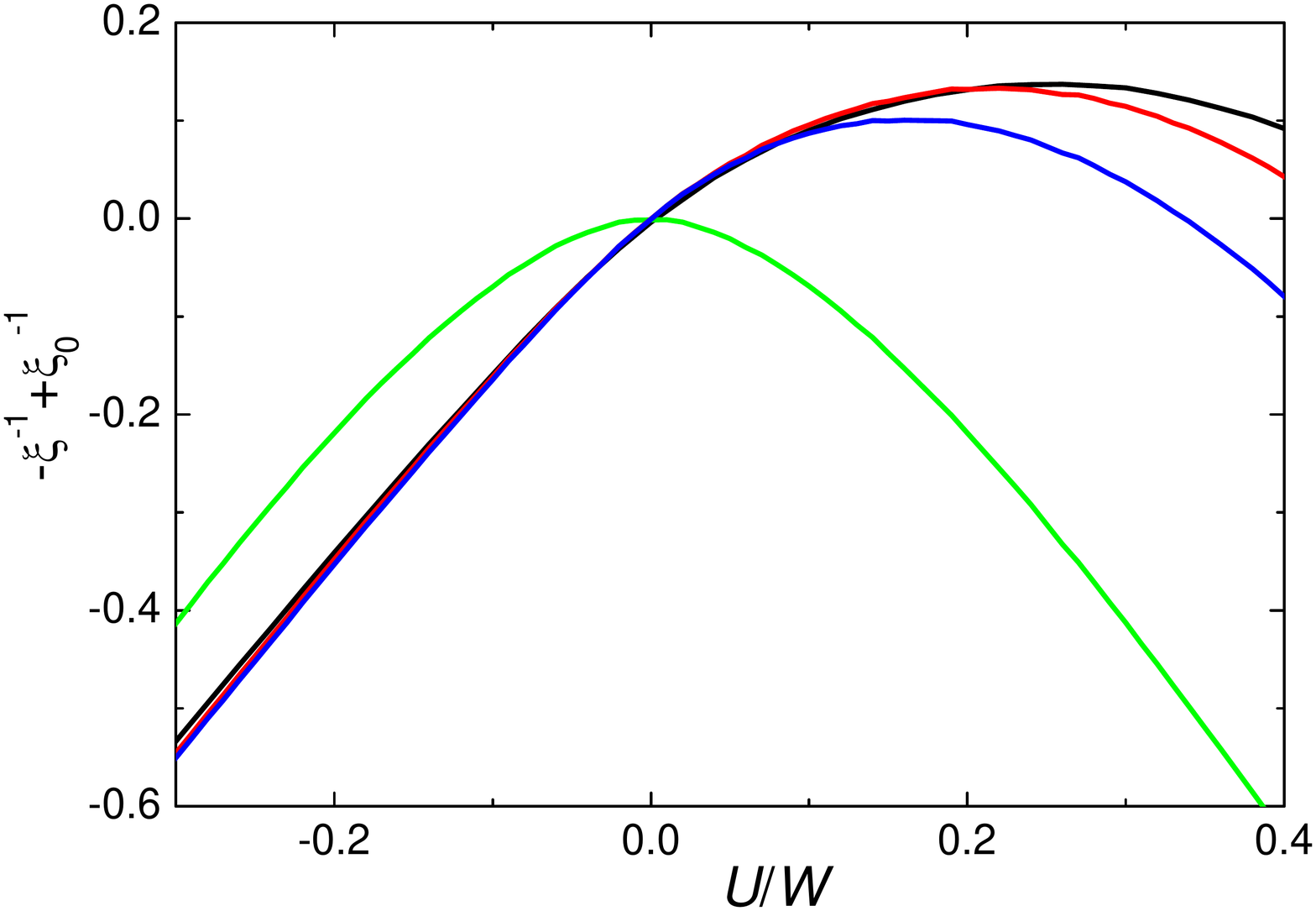}
\caption{Change in the inverse localization length with respect to the noninteracting case as a function of $U/W$, for average half occupation, for our perturbative procedure (red curve) and for our approximation (blue curve).
\label{approx}}
\end{figure}

In Fig.\ \ref{approx} we compare $-\xi^{-1}+\xi_0^{-1}$ as a function of $U/W$ for our perturbation procedure (black curve) and for the present approximation (green curve). The discrepancy between both results is quite large.

It is easy to improve the estimate (\ref{first}) drastically by analogy with the following exact result for the one-particle 
GF in 1D \cite{aronov}
\begin{equation}
G(x,x')=\frac{G(x,x'') G(x'',x')}{G(x'',x'')}.
\end{equation}
Based on this result, we propose as a second-order approximation
\begin{equation}
G(1,L;\omega)\approx G(1,2;\omega) \prod_{i=2}^{L-1} \frac{G(i,i+1;\omega)}{G(i,i;\omega)}.
\label{second}
\end{equation}
The blue curve in Fig.\ \ref{approx} corresponds to this approximation and we note that represents a big improvement with respect to (\ref{first}).

We note that in the places where the GF factorizes, as deduced in Sec.\ \ref{factor}, Eq.\ (\ref{second}) is exact (within our forward-scattering path approximation).

The previous scheme can be generalized to nth order
\begin{equation}
G(1,L;\omega)\approx G(1,n;\omega) \prod_{i=2}^{L-n} \frac{G(i,i+n-1;\omega)}{G(i,i+n-2;\omega)}.
\label{general}
\end{equation}

The results for $n=2$ are also shown in Fig.\ \ref{approx} (red curve). They already constitute and excellent approximation to our GF.

These results show that a factorized expression may represent properly the GF in all cases and approaches the exact result even for $U>W/2$, when the ground state corresponds to an alternated sequence. We note that one can expect factorization to become exponentially accurate for $U<W/2$ and in all cases if occupation is different from $1/2$

\section{Discusion}
\label{dis}

We have investigated the single-particle retarded GF in the strongly localized limit for interacting fermionic systems and we have obtained a diagrammatic representation in terms of forward-scattering paths. It is also possible to proceed in the same way for bosonic systems. We have also shown that the GF factorizes when the system can be separated into two non interacting parts. This theorem is of major importance because, as we have shown, in 1D with nearest neighbors interactions this factorization applies when two adjacent 
sites are both occupied or both empty. Similarly we can always expect factorization of the GF for more general short-range interactions. This fact leads to relevant consequences. According to single-parameter scaling we can expect the conductance distribution to be universal and we can claim that in the strongly localized limit, due to this factorization, interacting systems present a log-normal distribution.
We can also conclude that both mean and variance of $\log g$ grow linearly with $L$. We note that a dependence Var$(\log g) \propto L^{(2/3)}$ was claimed for the same model in the localized regime \cite{ScSc98}. According to our results this exponent should be 1 in the strongly-localized limit, when $L\rightarrow \infty$, and we believe that the exponent 2/3 was obtained without reaching the proper asymptotic behavior. In summary, we have shown that the conductance distribution for interacting systems is similar to the distribution for non interacting systems, and thus share the same universality class in the single parameter scaling limit.

A similar argument can be applied to 2D systems as well. In this case the 2D GF between points ${\bf r}_1$ and ${\bf r}_2$ is the sum of all 1D GF calculated along all possible directed paths connecting the two points.  Factorization into local contributions in real space is guaranteed at least for occupation far from $1/2$.

A possibility that could invalidate the previous conclusions 
would be a non perturbative character of the problem. If the system has a well defined localization length, in the renormalization group sense, we must expect a flow into the fixed point $\xi/L\rightarrow 0$ and our arguments should be valid. But it might happen that adding
any amount of interaction, the localization length could become not well defined. According to the many-body localization theory \cite{Altshuler} this is not the case near $T=0$, but it could occur at higher temperatures.  
 
Our result can be used to study systems in which tunnelling of  particles between different regions is weak.
For instance, experiments involving Josephson junctions arrays work often in a regime in which the Josephson energies, which control Cooper pair tunnel through the junctions, are much smaller than charging energies,  which control potential contribution to the energy due to an excess of Cooper pairs. Furthermore, the capacitative coupling between different superconducting islands are, in many occasions, very small or negligible\ \cite{BeIo12}. In this case, our result can be useful to factorize the total GF of Cooper pairs using contributions coming from different superconducting islands.

The main motivation to use the forward-scattering approximation was to  state in the simplest way the universality of the problem. But we have shown that this approximation could give good quantitative estimations of $\xi$ for large disorder. We believe that quantitative 
agreement would greatly improve after a proper treatment of the resonances of the problem.

In the limit of strong disorder and when the initial state is the ground state, there are no many resonances, i.e., factors larger than 1 in our path expansion. In order to extend the method to highly excited states, for which there are an exponentially large number of other states with roughly the same energy, one will have to handle resonances properly. Here we just mention a preliminary analysis of how to take resonances into account. 

The first type of resonance occurs when a site has an energy very close to the chemical potential, provided we are interested in the Green function at $\omega=\mu$.
In the noninteracting case, we just can consider that the inclusion of higher order terms in $t$ will renormalize the corresponding factor, which will never be larger than unity.
The previous intuitive result can be worked out exactly for a few sites. Let us consider just two sites.
The exact Green function is
\begin{equation}
G(1,2;\omega)=\frac{-t}{(\omega-\epsilon_1)(\omega-\epsilon_2)-t^2}\,.
\label{res1}\end{equation}
If $\omega\approx\epsilon_1$, the first term in the denominator is likely to be smaller than the second term
and could be neglected, but if $|\omega-\epsilon_2|\gg t$ this is not the case. Instead, it is better to approximate (\ref{res1}) by $1/(\omega-\epsilon_2)$, which always works fine.
When a site energy is close to the chemical potential, there is no hopping penalty in going through it.

When a site energy is close to $\omega$ in an interacting system,  we can keep  the terms containing large contributions due to the resonance and neglect the rest of terms in the sum over all paths. For example, for our previous two site problem, if we have $\omega\approx\epsilon_1+U$ we only keep the term
\begin{equation}
\frac{t}{\epsilon_1+U-\epsilon_2}\frac{1}{-\omega+\epsilon_1+U}
\label{res2}\end{equation}
The second factor is substituted by the bound $1/t$ and the GF becomes (taking into account that
$\omega\approx\epsilon_1+U$)
\begin{equation}
G(1,2;\omega)\approx\frac{1}{\omega-\epsilon_2}
\label{res3}\end{equation}
If the resonant site is the second one, which is empty in our example, $\omega\approx\epsilon_2+U$ and we arrive at
\begin{equation}
G(1,2;\omega)\approx\frac{1}{-\omega-\epsilon_1+2U}
\label{res4}\end{equation}
The general rule is the following. If a site resonates, there is no contribution from this site and the other sites see this site as always occupied if it was originally empty and, the other way around, always empty if it was originally occupied.
The presence of a resonant site does not invalidate the  factorization procedure. 

As we have already mentioned, our procedure also applies to bosons. One just have to take into account the proper commutation relations,
with the corresponding sign changes between different paths and possibility that any site is occupied by any integer number of bosons. The one-particle result is reproduced when the initial state is the vacuum state.
The GF for interacting bosons is most interesting for excited states and is the aim of future work.

We would like to remark that the factorization of the GF, proved in section \ref{factor}, is of very general applicability and, in particular, can be very useful in heterogeneous systems, as granular metals or polymer systems, where the entities composing the material are linked by small hopping and interaction energies, much smaller than the other typical energies involved in the transport process.

\appendix
\section{Three sites}
\label{appa}

Let us consider a three sites system, where we can make all calculations explicitly.
Site energies are $\epsilon_1$, $\epsilon_2$ and $\epsilon_3$ and we assume that the ground state, at zero order in $t$, corresponds to a single particle located at site 2.  We want to calculate the single particle GF between the two sites at the extremes, 1 and 3, $G(1,3;\omega)$

For the noninteracting case the result is independent of the occupancy of the ground state (apart from a possible sign). We can calculate it by propagating a particle in a vacuum state. Then, in the strongly localized regime there is a unique minimum path which consists in  going from site 1 to 2 and from 2 to 3:
\begin{equation}
G(1,3;\omega)=\frac{1}{\omega-\epsilon_1}t\frac{1}{\omega-\epsilon_2}t\frac{1}{\omega-\epsilon_3}\,.
\label{ap1}\end{equation}

We now want to reproduce this result, in the many-body formulation, by summing over all possible paths connecting the initial and the final
configurations according to the rules described in section \ref{paths}.
\begin{figure}
\includegraphics[width=.4\textwidth]{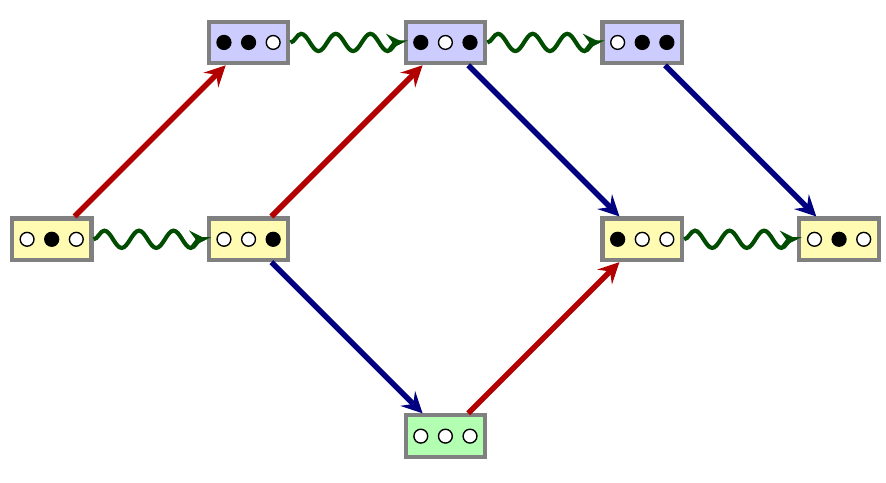}
\caption{Scheme of the configurations involved in the calculation of the GF for a system with three sites and an electron, originally placed at site in the middle. Each rectangle corresponds to a configuration. Red arrows mark the insertion
of an electron at site 1, while blue arrows signal the extraction of a particle at site 3, and wavy arrows propagation by expansion in the hopping $t$.
\label{fig5}}
\end{figure}
Fig.\ \ref{fig5} shows an scheme of the configurations and the transitions  involved in the calculation of the GF for a system with three sites and a particle, originally placed at site 2. Each rectangle corresponds to a configuration. The configurations in the middle row are one-electron configurations, while those on top have two electrons and the one at the bottom corresponds to the vacuum state.
Red (blue) arrows mark the insertion (extraction) of a particle at site 1 (3), while  wavy arrows refer to propagation by the hopping term, contributing with a factor $t$. The GF $G(1,3;\omega)$ is the sum of the contributions from all possible paths. In all paths particles move only to the right, forward scattering approximation, since we are in the strongly localized regime and want to minimize hopping.

The path corresponding to the propagation of a hole from right to left visits configurations $|2\rangle \rightarrow |3\rangle \rightarrow | 0\rangle \rightarrow  |1\rangle \rightarrow  |2\rangle $, 
where $|0\rangle $ refers to the vacuum state and $|i\rangle $, with $i>0$ to the one-particle configuration with the particle at site $i$. That is, in this path the particle is initially at site 2, then moves to 3
where is removed and a new particle is inserted in 1, which finally moves to 2 to complete the path. Its contribution to the GF is
\begin{equation}
\frac{t^2}{(\epsilon_2-\epsilon_3)(-\omega+\epsilon_2)(\epsilon_2-\epsilon_1)}
\label{ap9}\end{equation}
where the two $t$ factors come from the two wavy lines and the three factors in the denominator come from configurations $|3\rangle $, $|0\rangle $ and $|1\rangle $, respectively. 

Propagation of a particle to the right involves four paths. The contribution of subpaths 
$|2\rangle \rightarrow |1,2\rangle \rightarrow |1,3\rangle $ (where $|i,j\rangle $ is a two-particle state with particles at $i$ and $j$) and 
$|2\rangle \rightarrow |3\rangle \rightarrow |1,3\rangle $ is
\begin{eqnarray}
\frac{t}{(\omega-\epsilon_1)(\omega+\epsilon_2-\epsilon_1-\epsilon_3)}
&+&\frac{t}{(\epsilon_2-\epsilon_3)(\omega+\epsilon_2-\epsilon_1-\epsilon_3)}\nonumber\\
&=&\frac{t}{(\omega-\epsilon_1)(\epsilon_2-\epsilon_3)}
\label{ap10}\end{eqnarray}
To complete the four paths for particle propagation, the previous factor has to be multiplied by the contribution from configurations 
$|2,3\rangle $ and  $|1\rangle $
\begin{equation}
\frac{t}{\omega-\epsilon_3}+\frac{t}{\epsilon_2-\epsilon_1}
\label{ap11}\end{equation}
resulting in
\begin{equation}
\frac{t^2}{(\omega-\epsilon_1)(\epsilon_2-\epsilon_3)}\left[\frac{t}{\omega-\epsilon_3}+\frac{t}{\epsilon_2-\epsilon_1}\right]
\label{ap6}\end{equation}
Adding this term and (\ref{ap9}) we recover (\ref{ap1}) for the noninteracting case.

The advantage of our formulation is that it can be easily generalized to the interacting case. In our example, we just have to add $U_1$ to the energy of configurations $|1,2\rangle $  and $U_3$  to the energy of $|2,3\rangle $ . The final result is
\begin{equation}
G(1,3)=\frac{1-\frac{U_1}{\epsilon_2-\epsilon_1}-\frac{U_3}{\epsilon_2
-\epsilon_3}+
\frac{U_1U_3}{w+\epsilon_2-\epsilon_1-\epsilon_3}(\frac{1}{
\epsilon_2-\epsilon_1}+\frac{1}
{\epsilon_2-\epsilon_3})}
{(w-\epsilon_1-U_1)(w-\epsilon_2)(w-\epsilon_3-U_3)}.\label{Ga}
\end{equation}
We kept $U_1$ and $U_3$ independent in order to show that if $U_3=0$ the GF factorizes,
$G(1,3)=G(1,2)tG(3,3)$.

We can check the validity of this result from the spectral representation of the GF,  
Eq.\ (\ref{omega}), which for this small system can be easily calculated.
The initial state is
\begin{equation}
| \Psi_2\rangle =  |2\rangle + \frac{t}{\epsilon_2-\epsilon_1}|1\rangle+  \frac{t}{\epsilon_2-\epsilon_3}|3\rangle\,.
\label{ap2}\end{equation}
Let us evaluate first the second term on the RHS of  Eq.\ (\ref{omega}), corresponding to the hole propagation. As we initially have one single particle, the only state with one particle less is the vacuum $|0\rangle$, and
we need
\begin{align}
\langle\Psi_2|c_3^\dagger| 0\rangle &= \frac{t}{\epsilon_2-\epsilon_3}\nonumber\\
\langle 0|c_1| \Psi_2\rangle &= \frac{t}{\epsilon_2-\epsilon_1}\,.\nonumber
\label{ap7}\end{align}
The hole contribution to the GF is then equal to Eq.\ (\ref{ap9}) changed sign.

 We evaluate now the first term on the RHS of  Eq.\ (\ref{omega}), corresponding to particle propagation. An adequate basis for two-particle states are the three antisymmetric two-particle wave functions, whose relevant terms for the matrix elements up to second order in $t$ are 
\begin{widetext}
\begin{eqnarray}
|\Psi_{1,2}\rangle &=&  |1,2\rangle + \frac{t}{\epsilon_2-\epsilon_3+U}|1,3\rangle+  
\frac{t^2}{(\epsilon_2-\epsilon_3+U)(\epsilon_1-\epsilon_3)}|2,3\rangle\nonumber\\
|\Psi_{1,3}\rangle &=&  |1,3\rangle + \frac{t}{\epsilon_3-\epsilon_2-U}|1,2\rangle+  
\frac{t}{\epsilon_1-\epsilon_2-U}|2,3\rangle\\
|\Psi_{2,3}\rangle &= & |2,3\rangle + \frac{t}{\epsilon_2-\epsilon_1+U}|1,3\rangle+  
\frac{t^2}{(\epsilon_2-\epsilon_1+U)(\epsilon_3-\epsilon_1)}|1,2\rangle\nonumber
\label{ap3}\end{eqnarray}
\end{widetext}
To lowest order, the state $|\Psi_{1,3}\rangle$ does not contribute and the contribution
from the other two states is
\begin{eqnarray}
\left[\frac{t}{\omega-\epsilon_1-U}+\frac{t}{\epsilon_2-\epsilon_3}\right]
\frac{1}{\omega+\epsilon_2-\epsilon_1-\epsilon_3}\times\nonumber\\
\left[\frac{t}{\omega-\epsilon_3-U}+\frac{t}{\epsilon_2-\epsilon_1}\right]
\label{ap5}\end{eqnarray}
Adding (\ref{ap5}) and (\ref{ap9}), we arrive at the final result for the GF (\ref{Ga}).
If $U=0$, Eq.\ (\ref{ap5}) reduces to the result for the noninteracting case, Eq.\ (\ref{ap6}).

\begin{acknowledgments}
The work was supported by the
Spanish DGI and FEDER Grant No. FIS2012-3820 and by a grant from ARO (W911NF-13-1-043).
MO thanks KITP for hospitality.
\end{acknowledgments}

\end{document}